\documentstyle[jkas]{article}

\beginpage{135}
\endpage{139}
\year{2010}\volume{43}\month{August}

\runningauthor {YI \& KIM} 
\runningtitle{IMPACT OF THE LOW SOLAR ABUNDANCE}

\month{August} \year{2010} \volume{43}
\beginpage{135}\endpage{139}
\date{Received December 1, 2009; Accepted January 17, 2010}

\newcommand{\dbv}{$\Delta\,(B-V)\,$}                   

\begin{document}
\title{IMPACT OF THE LOW SOLAR ABUNDANCE ON THE AGES OF GLOBULAR CLUSTERS}
\author{Sukyoung K. Yi and Yong~-Cheol~Kim}
\address{Department of Astronomy, Yonsei University, Seoul 120-749, Korea\\
 {\it E-mail : yi@yonsei.ac.kr, yckim@yonsei.ac.kr}}

\address{\normalsize{\it (Received April 1, 2010; Accepted July 10, 2010)}}
\offprints{Y.~-C.~Kim}
\abstract{
We present the result of our investigation on the impact of the
low Solar abundance of Asplund and collaborators (2004)
on the derived ages for the oldest star clusters based on isochrone fittings.
We have constructed new stellar models and corresponding
isochrones using this new solar mixture with a proper Solar calibration.
We have found that the use of the \citet{asp04}
metallicity causes the typical
ages for old globular clusters in the Milky Way to be increased roughly
by 10\%. Although this may appear small, it has a significant impact on
the interpretation for the formation epoch of Milky Way globular clusters.
The \citet{asp04} abundance may not necessarily threaten the current
concordance cosmology but would suggest that Milky Way globular clusters
formed before the reionization and before the main galaxy body starts to
build up. This is in contrast to the current understanding on the galaxy
formation.}


\maketitle

\section{INTRODUCTION}
The Solar mixture of \cite{asp04} suggests a surprisingly-low
metallicity for the Solar atmosphere, $Z/X=0.0165$, compared to
the previous estimates (e.g., $Z/X=0.0231$ \citet{gre98}). 
Their mixture has been challenged by a few studies since
\citep[e.g.,][]{ba04,lan07,cs08}.
Although recently they have revised their mixture whose $Z/X=0.0181$
\citep{asp09}, it is still uncomfortably lower than the values regarded
in general.
It obviously would have profound impacts on many subfields of astronomy
if found to be true.

In this paper we present the result of our
investigation on its impact on the age derivation for globular clusters
based on isochrone fittings. 
We have adopted \citet{asp04} mixture which is more extreme. 
Old globular clusters in the Milky Way galaxy
are often considered the oldest stellar objects in the universe and thus
place an important constraint to our cosmological paradigm.

Stellar isochrones are the most powerful and thus widely-used tools
for deriving the ages of globular clusters.
This is possible mainly because globular clusters are considered
``simple'' stellar populations that are composed
of stars of effectively the same age and chemical composition.
The validity of the ``simple'' population assumption is challenged
by a number of recent studies that found multiple stellar populations
in some globular clusters \citep[e.g.,][]{bed04, nor04, pio05}, yet globular
clusters are still the simplest populations we know of.

Numerous improvements in microphysics, such as opacities and equations
of states, have polished stellar models steadily, and we now
boast knowing the ages of globular clusters and thus the lower limit
on the age of the universe with unprecedented accuracy
(e.g., for globular cluster ages \citet{mar09,mar10};
for stellar ages \citet{sod10}).
Together with the recent advances in the cosmological understanding,
notably from cosmic microwave background radiation and dark matter/energy
studies this has finally led us to achieve a concordance model
in the big bang paradigm.
The basic teachings of the concordance model include the age of the universe,
roughly 13.7~Gyr \citep{kom10}, the existence of reionization at
roughly 1 billion years after the big bang \citep{bec01,dun09,kom10},
and the formation of galaxies after reionization.
First stars, deemed responsible for reionization, were the only
astronomical objects that are considered to have formed before reionization,
and their traces are found only in the metal abundance in population
II stars. In conclusion, no stellar objects are expected in the
concordance cosmology to have ages greater than 13 billion years.
In this Letter we show how the recent Solar mixture affects the derived ages
of globular clusters and such a concordance prediction.

\begin{figure*}[t]
\centering
\epsfxsize=14cm 
\epsfbox{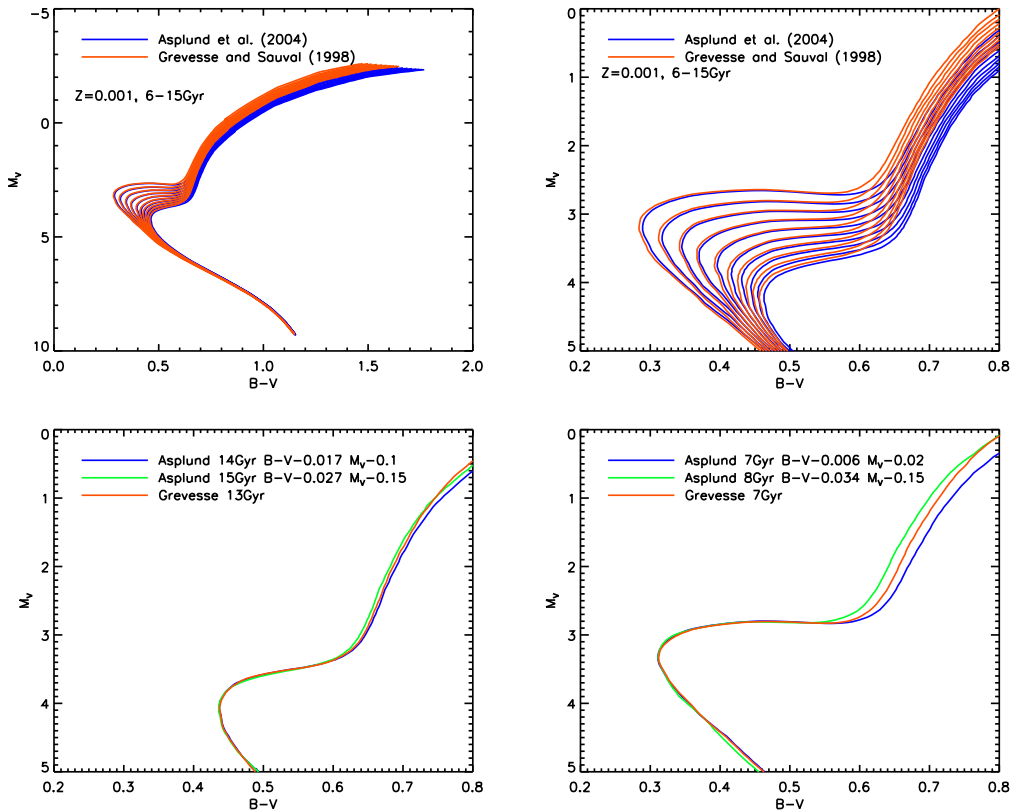} 
\caption{
The isochrones based on the previous Grevesse and Sauval (1998) mixture
and the recent \citet{asp04} mixture.
(top left) Isochrones for $X=0.767$, $Z=0.001$, $[\alpha/Fe]=0.3$, and
ages = 6 through 15 Gyr with 1Gyr spacing.
(top right) Zoom-in figure of the top left panel.
(bottom left) The 14Gyr isochrone based on the \citet{asp04} mixture
closely reproduces the 13Gyr isochrone based on the Grevesse \& Sauval
mixture.
(bottom right) The 8Gyr isochrone based on the \citet{asp04} mixture
closely reproduces the 7Gyr isochrone based on the Grevesse \& Sauval
mixture.  }
\label{fig1}
\end{figure*}

\section{STELLAR MODEL CONSTRUCTION}
We have used the Yale stellar evolution code, $YREC$, to construct stellar
models.
For the reference model, we have adopted the \citet{gre98} mixture for the
Sun.
And for our comparison model, we use the recent \citet{asp04} mixture.
Other microphysics prescriptions were kept the same.
The input physics details can be found in \citet{yi01} and \citet{kim02}.
In order to construct the isochrones for the age range
6 -- 15\,Gyr that extend to the tip of the RGB, we have constructed stellar
models of mass 0.4 -- 1.2 $M_\odot$ with 0.05 $M_\odot$ spacings.

Our stellar models, regardless of the choice of the Solar mixture,
have been calibrated against the sun, following the standard practise
as described in \citet{yi01}.
The mixing length parameters of $l/H_{p} = 1.811$ and 1.742 have been
found to match the Solar properties for the \citet{gre98} and \citet{asp04}
mixtures, respectively. A lower value of metallicity causes lower opacities
in the Solar atmosphere which make it easier for photons to escape the
Sun. This causes a reduction in the stellar radius. A smaller value of
mixing length parameter counteracts this effect.

\section{IMPACT ON THE AGE ESTIMATES}
Isochrones are used for age derivations mainly in two approaches.
One is to use the width between the main-sequence turn off and
the bottom of the red giant branch, a.k.a. the $\Delta (B-V)$ method.
The other is to use the height of the horizontal branch measured from
the main-sequence turn off.
Reviews are available on these techniques \citep[e.g.,][]{van96,sar97}.

\subsection{$\Delta (B-V)$ Method}
Fig.~\ref{fig1} shows the result of our comparison for a typical
metal-poor populations ($Z=0.001$) using the $\Delta (B-V)$ method.
The top panels show the isochrones for the two Solar mixtures.
A cursory inspection might suggest little impact.
But the lower panels suggest otherwise.
The lower panels compare the two sets of isochrones at two different
ages, 13Gyr (left) and 7Gyr (right).
The width between the main-sequence turn off and the bottom of the
sub giant branch (hereafter, \dbv) has become larger as we switched
the mixture from \citet{gre98} to \citet{asp04}.
The use of \citet{asp04} mixture suggests a larger age by 7--8 percent
for this age range and metallicity.
Similar exercises for other metallicities yield different amount
of age increase.
Isochrone ages increase roughly by 5--8, 7--8, and 14--15 percent as
we change the metallicity from $Z=0.0001$, through 0.001 and to the
Solar metallicity.
That is, the impact of the mixture on age estimates is greater for higher
metallicities, as expected.

\begin{figure}[!t]
\centering
\epsfxsize=8cm 
\epsfbox{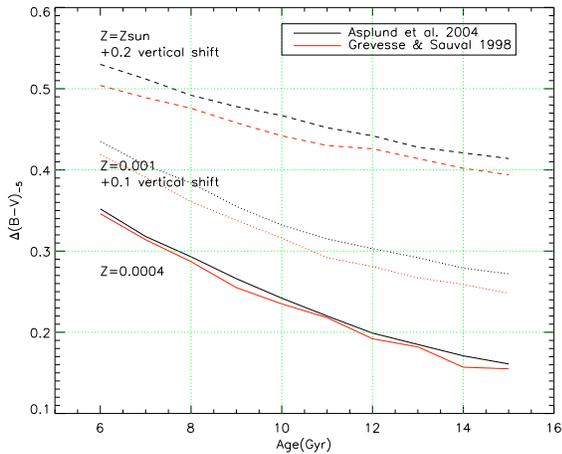} 
\caption{
The $\Delta(B-V)$ as a function of age for three different metallicities 
and two different mixtures. For visual clarity, vertical shifts were 
applied to the two more metal-rich models. For given values 
of $\Delta (B-V)$ the use of Asplund et al. (2004)
mixture suggests systematically larger ages.  }
\label{fig2}
\end{figure}

The evolution of $\Delta (B-V)$ with time and the impact of the change of
the Solar mixture are more clearly illustrated in Fig.~\ref{fig2}.
We measure $\Delta (B-V)$ as a horizontal distance in the color-magnitude
diagram between the main-sequence turn off (the bluest point) and
the base of the subgiant branch. We made an arbitrary decision for the location
of the base of the subgiant branch, that is, the place where the slope of
$\delta M_{\rm V} / \delta (B-V) =-5$. The change of this criterion does not
make a significant impact on the result.

\begin{figure}[!t]
\centering 
\epsfxsize=8cm 
\epsfbox{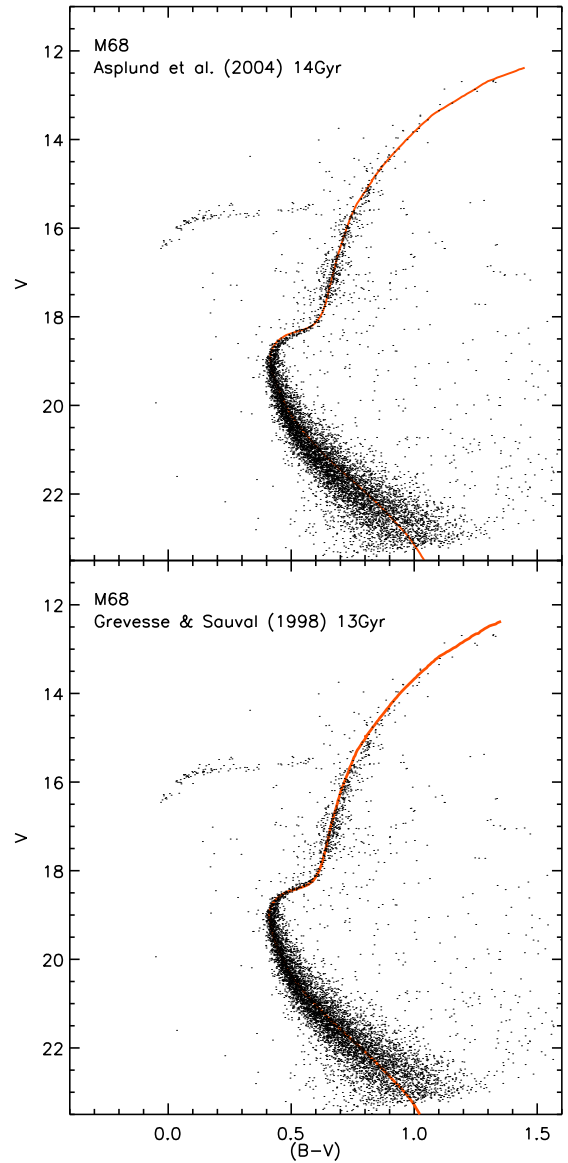}
\caption{
Sample isochrone fits to the data of M68 using the \citet{asp04} mixture
(top) or the Grevesse \& Sauval (1998) mixture (bottom).
Cluster properties were kept the same:
$[\rm Fe/ \rm H]=-2.10$, $E(B-V)=0.02$, $(m-M)_V=15.17$.}
\label{fig3}
\end{figure}

Fig.~\ref{fig3} shows the isochrone fits to the observed data for
the globular cluster M68. The data are from Walker (1994; private comm.).
The estimates of [Fe/H] and reddening are from Harris (1996), and
[$\alpha$/Fe]=0.3
has been adopted following the halo observations.
It shows that one can achieve the same quality isochrone fits
regardless of the choice of the mixture.

\subsection{$\Delta M_V (\rm ZAHB-TO)$ Method}

\begin{figure}[!t]
\centering 
\epsfxsize=8cm 
\epsfbox{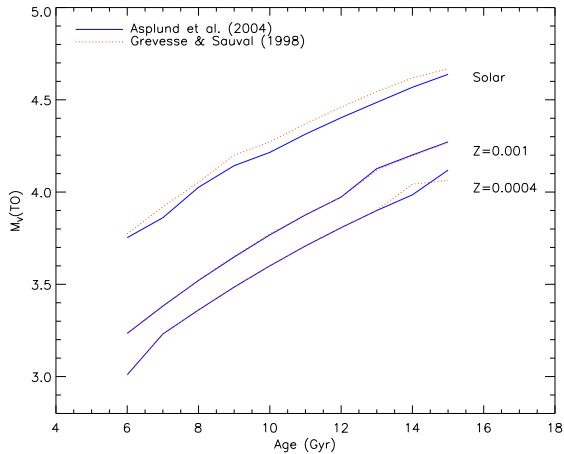}
\caption{The magnitude difference at the main-sequence turn off
with respect to age, based on the two different mixtures.}
\label{fig4}
\end{figure}

We do not have the horizontal branch stellar models yet for the
recent mixture. So in this work we compare just the main-sequence
brightness when the two different mixtures are used.
Fig.~\ref{fig4} shows the main-sequence turn-off luminosity
for the age range of 6 -- 15Gyr.
For sub-solar metallicities, the brightness difference is negligible.
The difference becomes notable for the solar metallicity, obviously
because the two mixtures assume different metallicities for the Sun.
For the typical metallicity range for metal-poor globular clusters,
that is, $Z=0.0001$ -- 0.001, the choice of the Solar mixture
does not appear to affect much the main-sequence brightness and thus
$\Delta M_V (\rm ZAHB-TO)$-based isochrone ages.
\citet{del06} have explored this effect and concluded that the maximum
variation in the age estimates is of the order of 10\% but only in young star
clusters.
Our investigation on this vertical method is incomplete and a
more thorough analysis would require horizontal-branch
stellar models in addition.

\section{DISCUSSION}
The arguably more robust age derivation method, the $\Delta (B-V)$ method,
is somewhat sensitive to the choice of Solar mixture.
The ages of old metal-poor globular clusters would be roughly 5--10\%
larger if the recent \citet{asp04} mixture is adopted.
The magnitude of the sensitivity may appear small but significant from the
perspective of galaxy formation.
The typical age of old metal-poor globular clusters would become substantially
larger than 13Gyr, which is very close to the current estimate for the
age of the universe based on the concordance model.
This may have an important implication to the current understanding on the
formation of the first and second generations of stars.
If many of the Milky Way globular clusters indeed formed so close
to the age of the universe, it could  mean that they formed even before
the reionization.
This would be theoretically implausible considering that these
clusters already have non-zero metallicities and would not have had
enough time to be chemically enriched.
In this respect, the recent \citet{asp04} mixture appears incompatible with the
concordance picture of cosmology and galaxy formation.

\acknowledgments{
We thank J.~W.~Ferguson for providing us with 
the low temperature opacity tables
for the Asplund mixture.
We thank Alistair Walker for making the M68 CMD 
data available to us. 
We are also grateful to Do-Gyun Kim and Sang-Il Han for assists
in the figure generation.
This work was supported by the Korea Research Foundation Grant
funded by the Korean government (KRF-C00156).  }



\begin{thebibliography}{}

\bibitem[Asplund \etal(2004)]{asp04}
Asplund, M., Grevesse, N., Sauval, A. J., Allende Prieto, C., \&
 Kiselman, D.\ 2004, Line formation in solar granulation. IV.
 [O I], O I and OH lines and the photospheric O abundance,
\aaps, 417, 751

\bibitem[Asplund \etal(2009)]{asp09}
Asplund, M., Grevesse, N., Sauval, A. J., \& Scott, P.\ 2009, 
The Chemical Composition of the Sun, ARA\&A, 47, 481

\bibitem[Basu \& Antia(2004)]{ba04}
Basu, S. \& Antia, H.M.\ 2004, Constraining Solar Abundances 
Using Helioseismology,
\apj, 606, 85

\bibitem[Becker \etal(2001)]{bec01}
Becker, R. H. et al.\ 2001, 
Evidence for Reionization at $z~6$: Detection of 
a Gunn-Peterson Trough in a $z=6.28$ Quasar,
\aj, 122, 2850

\bibitem[Bedin \etal(2005)]{bed04}
Bedin, L. R., Piotto, G., Anderson, J., Cassisi, S., King, I. R., Momany, Y.,
  \& Carraro, G.\ 2004, 
Centauri: The Population Puzzle Goes Deeper,
\apj, 605, L125

\bibitem[Centeno \& Socas-Navarro(2008)]{cs08}
Centeno, R. \& Socas-Navarro, H.\ 2008, 
A New Approach to the Solar Oxygen Abundance Problem,
\apj, 682, 61

\bibitem[Degl'Innocenti \etal(2006)]{del06} 
Degl'Innocenti, S., Prada Moroni, P.G.,
\& Ricci, B.\ 2006, 
The heavy elements mixture and the stellar cluster age,
Astrophys. Space Sci, 305, 67

\bibitem[Dunkley \etal(2009)]{dun09}
Dunkley, J. et al.\ 2009, 
Five-Year Wilkinson Microwave Anisotropy Probe Observations: Likelihoods and
Parameters from the WMAP Data,
\apjs, 180, 306

\bibitem[Ferreras \& Silk(2001)]{fs01}
Ferreras, I. \& Silk, J.\ 2001, 
A Backwards Approach to the Formation of Disk Galaxies. I. Stellar and Gas
Content,
\apj, 557, 165

\bibitem[Grevesse \& Sauval(1998)]{gre98}
Grevesse, N. \& Sauval, A. J.\ 1998, 
Standard Solar Composition,
Space Sci. Rev. 85, 161

\bibitem[Harris(1996)]{h96}
Harris, W. E.\ 1996, 
A Catalog of Parameters for Globular Clusters in the Milky Way,
\aj, 112, 1487

\bibitem[Kim \etal(2002)]{kim02}
Kim, Y.-C., Demarque, P., Yi, S., \& Alexander, D.\ 2002, 
The Y$^2$ Isochrones $\alpha$-Element Enhanced Mixtures,
\apjs, 143, 499

\bibitem[Komatsu \etal(2010)]{kom10}
Komatsu, E. et al. 2010, 
Seven-Year Wilkinson Microwave Anisotropy Probe (WMAP) Observations:
Cosmological Interpretation,
\apjs, submitted, arXiv:1001.4538v1

\bibitem[Landi \etal(2007)]{lan07}
Landi, E., Feldman, U., \& Doschek, G. A.\ 2007, 
Neon and Oxygen Absolute Abundances in the Solar Corona,
\apj, 659, 743

\bibitem[Marin-Franch \etal(2009)]{mar09}
 Marin-Franch, A., et al.\  2009,
 The ACS Survey of Galactic Globular Clusters. VII. Relative Ages,
 \apj, 694, 1498 

\bibitem[Marin-Franch \etal(2010)]{mar10}
 Marin-Franch, A., Cassisi, S., Aparicio, A., \& Pietrinferni, A.\  2010,
The Impact of Enhanced He and CNONa Abundances on Globular Cluster 
Relative Age-Dating Methods,
\apj, 714, 1072

\bibitem[Norris(2004)]{nor04}
Norris, J. E.\ 2004, 
The Helium Abundances $\omega$ Centauri,
\apj, 612, 25

\bibitem[Piotto \etal(2005)]{pio05}
Piotto et al. 2005, 
Metallicities on the Double Main Sequence $\omega$ Centauri 
Imply Large Helium Enhancement,
\apj, 621, 777

\bibitem[Sarajedini \etal(1997)]{sar97}
Sarajedini, A., Chaboyer, B., \& Demarque, P.\ 1997, 
The Relative Ages of Galactic Globular Clusters,
\pasp, 109, 1321

\bibitem[Soderblom(2010)]{sod10}
 Soderblom, D. R.\ 2010, 
 The Ages of Stars,
ARA\&A, to be appeared, arXiv1003.6074v1 

\bibitem[VandenBerg \etal(1996)]{van96}
VandenBerg, D. A., Bolte, M., \& Stetson, P.B.\ 1996,
The Age of the Galactic Globular Cluster System,
ARA\&A, 34, 461

\bibitem[Yi \etal(2001)]{yi01}
Yi, S., Demarque, P., Kim, Y.-C., Lee, Y.-W., Ree, C.H., Lejeune, T.,
\& Barnes, S.\ 2001,
Toward Better Age Estimates for Stellar Populations: 
The Y$^2$ Isochrones for Solar Mixture,
\apjs, 136, 417

 
\end{thebibliography}
\end{document}